\documentclass[runningheads]{llncs}
\usepackage[T1]{fontenc}
\usepackage{amsmath,amsfonts}
\usepackage{algorithmic}
\usepackage{algorithm}
\usepackage{array}
\usepackage[caption=false,font=normalsize,labelfont=sf,textfont=sf]{subfig}
\usepackage{textcomp}
\usepackage{stfloats}
\usepackage{url}
\usepackage{verbatim}
\usepackage{graphicx}
\usepackage{cite}
\usepackage{ulem}
\usepackage{xcolor}
\definecolor{darkgreen}{rgb}{0,0.5,0}
\begin{document}
\title{Image Complexity-Aware Adaptive Retrieval for Efficient Vision-Language Models}
\titlerunning{Image Complexity-Aware Adaptive Retrieval}
\author{Mikel Williams-Lekuona \and Georgina Cosma}
\authorrunning{Williams-Lekuona and Cosma}
\institute{Computer Science, Loughborough University, Loughborough, UK\\
\email{M.Williams@lboro.ac.uk, G.Cosma@lboro.ac.uk}}
\maketitle              % typeset the header of the contribution
\begin{abstract}
	Vision transformers in vision-language models typically use the same amount of compute for every image, regardless of whether it is simple or complex. We propose ICAR (Image Complexity-Aware Retrieval), an adaptive computation approach that enables vision transformers to use less compute for simple images whilst processing complex images through their full network depth. The key challenge is maintaining cross-modal alignment: embeddings from different processing depths must remain compatible for text matching. ICAR solves this through dual-path training that produces compatible embeddings from both the early-exit and full-depth paths. This maintains compatibility between image representations and text embeddings in the same semantic space, whether an image exits early or processes fully. Unlike existing two-stage approaches that require expensive reranking, ICAR enables direct image-text matching without additional overhead. To determine how much compute to use, we develop ConvNeXt-IC, which treats image complexity assessment as a classification task. By applying modern classifier backbones rather than specialised architectures, ConvNeXt-IC achieves state-of-the-art performance, attaining a Pearson correlation coefficient of 0.959 with human labelling whilst delivering 4.4$\times$ faster complexity prediction. Evaluated on standard benchmarks augmented with real-world web data, ICAR achieves 20\% faster image encoding while maintaining category-level performance and 95\% of instance-level performance, enabling sustainable scaling of vision-language systems.

\end{abstract}
\section{Introduction}

\begin{figure}[t]
    \centering
    \includegraphics[width=\linewidth]{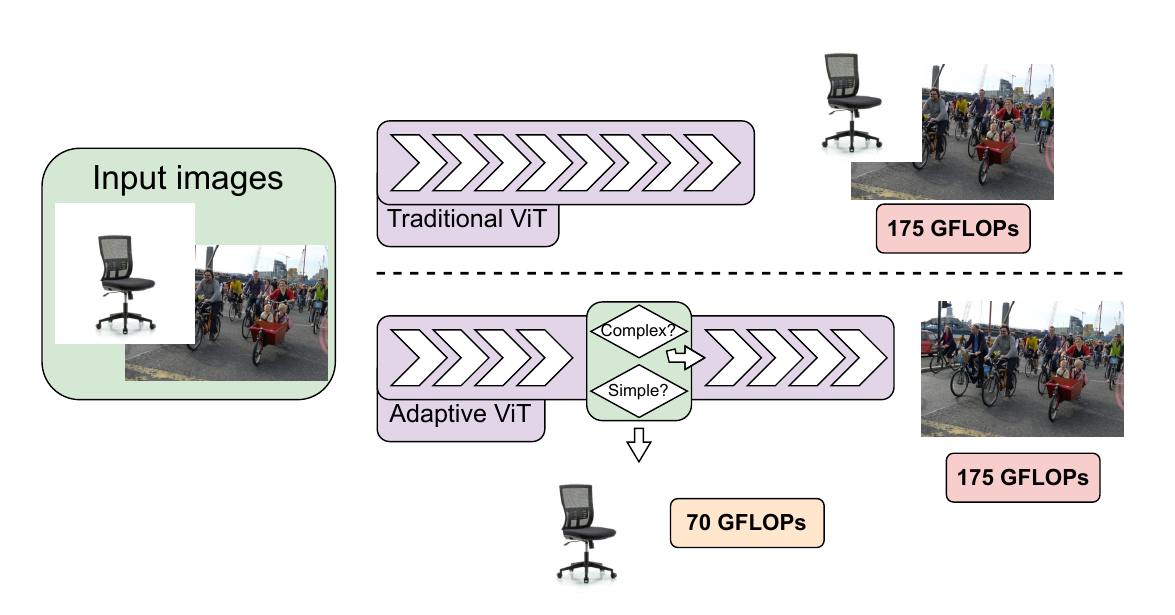}
    \caption{Adaptive computation for vision transformers. Traditional ViTs apply uniform processing to all images, while adaptive approaches route simple images through fewer layers, reducing computational cost.}
    \label{fig:adaptive_motivation}
\end{figure}

Many vision--language retrieval systems use Vision Transformers (ViT) to align images with text queries, enabling applications from web image search~\cite{radford2021learning} to e-commerce product matching~\cite{zhang2025research}. ViTs process images through multiple layers in sequence, progressively refining their representations~\cite{raghu2021vision}.
		
Standard ViT systems use a fixed-depth transformer stack and process all image patches uniformly, regardless of image complexity. For example, a 24-layer ViT-L/14 uses the same 175.33 GFLOPs~\cite{ilharco_gabriel_2021_5143773} (giga floating-point operations) of compute whether it's processing a simple product photo or a complex street scene. At web scale, where major platforms handle billions of photos and videos daily~\cite{GoogleIO2024}, this uniform processing can lead to substantial waste of computational resources. Adaptive computation can reduce this inefficiency by allowing simpler images to stop after fewer transformer layers, whilst complex images continue through the full transformer stack.
		
Adaptive computation for vision-language models faces a fundamental technical challenge in maintaining cross-modal alignment when embeddings are produced from different processing depths. While adaptive computation has been extensively studied for single-modality tasks \cite{teerapittayanon2016branchynet}, the adaptive vision-language field remains underexplored. Specifically, no existing work addresses adaptive computation for single-stage image--text retrieval where embeddings from different vision depths remain directly comparable to the text embedding.

This paper proposes Image Complexity-Aware Retrieval (ICAR), which uses image complexity to determine whether an image is routed through fewer transformer layers (early exit) or the full transformer stack (Figure~\ref{fig:adaptive_motivation}). ICAR introduces dual-path training that produces compatible embeddings from both early reduced compute and full processing paths. This enables simple images to use fewer layers whilst complex images process through the full network, all while maintaining cross-modal alignment through unified embedding spaces, eliminating the reranking overhead that existing approaches have.

To support these routing decisions, we develop ConvNeXt-IC (ConvNeXt-Image Complexity). While existing methods treat complexity assessment as a representation learning problem requiring specialised architectures, we instead recognise it as a classification task. We evaluate on both established academic benchmarks (IC9600) and real-world web data (LAION), demonstrating that our approach generalises beyond controlled settings. The following are the contributions of the paper:

\begin{itemize}
\item \textbf{Adaptive computation for single-stage image-text retrieval:} We introduce ICAR, which maintains cross-modal embedding compatibility across vision depths, enabling direct text matching without reranking overhead.

\item \textbf{Reconceptualising complexity assessment as classification:} We show that image complexity assessment is better suited as a classification task than representation learning, using modern backbones instead of specialised complexity assessment architectures.

\item \textbf{Efficiency--quality trade-off in retrieval:} We show that early exits can reduce image encoding compute with minimal impact on retrieval quality, enabled by dual-path training.
\end{itemize}

This paper is organised as follows: Section~\ref{sec:related} discusses related work in image complexity assessment and adaptive computation. Section~\ref{sec:complexity} presents our ConvNeXt-IC complexity detection method with benchmark evaluation on established benchmarks and real-world data. Section~\ref{sec:icar} details the ICAR architecture and dual-path training approach. Section~\ref{sec:results} presents experimental results and analysis of retrieval performance and efficiency gains. Finally, Section~\ref{sec:conclusion} concludes with key findings and future directions.

\section{Related Work}\label{sec:related}

\subsection{Image Complexity Assessment}

Image complexity assessment determines how difficult an image is for humans to understand or describe~\cite{snodgrass1980standardized}. Psychology research identified key factors: object count, spatial relationships, visual diversity, and background richness~\cite{olivia2004identifying,forsythe2009visual,purchase2012predicting}. Early computational approaches used hand-crafted features like entropy and compression ratios~\cite{stamps2002entropy,machado2015computerized,rosenholtz2007measuring}, but these worked only on small datasets.

Current deep learning methods treat complexity assessment as a representation learning problem. ICNet~\cite{feng2022ic9600} uses a dual-branch architecture combining detail and context pathways with spatial attention modules. ICCORN~\cite{guo2023image} applies ordinal regression to ICNet features, explicitly modelling the ordinal structure of complexity ratings. MICM~\cite{li2025unlocking} introduces motion-inspired assessment using image-to-video generation with hierarchical alignment losses. CLICv2~\cite{liu2025clicv2} uses self-supervised pretraining on complexity-related visual patterns.

Related image quality assessment methods explore similar perceptual challenges but focus on quality rather than complexity. HyperIQA~\cite{su2020blindly} employs content-adaptive hyper networks for quality assessment, while CLIPIQA~\cite{wang2022exploring} leverages vision-language representations from CLIP~\cite{radford2021learning}. TOPIQ~\cite{chen2024topiq} adopts a top-down framework modelling human visual perception characteristics.

These methods follow a common pattern: start with classification baselines, then add complexity-specific architectures for incremental improvements. However, they test only against outdated networks like AlexNet~\cite{krizhevsky2012imagenet} (2012) and ResNet~\cite{he2016deep} (2016), leaving a large gap unexplored. This raises an open question whether architectural specialisation provides genuine benefits, or whether stronger classification backbones would close the performance gap without the need for additional components. We address this gap using ConvNeXt-V2 (2023) to test whether modern classification architectures outperform specialised complexity assessment designs (Section~\ref{sec:complexity}).

\subsection{Adaptive Computation in Machine Learning}

\begin{figure*}[htbp]
    \centering
    \includegraphics[width=\linewidth]{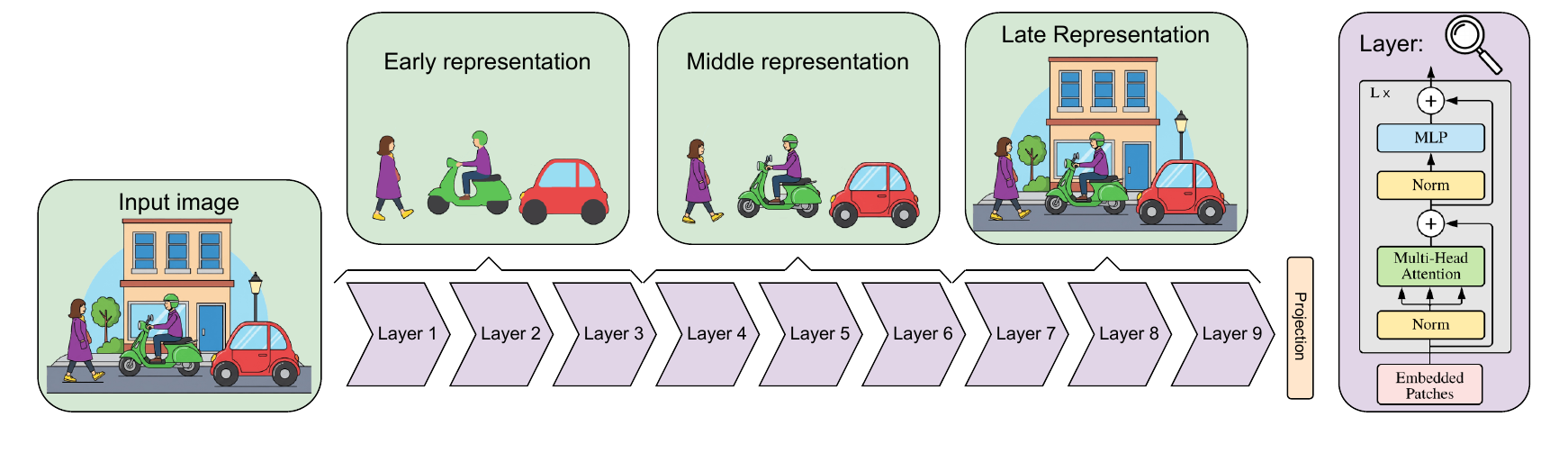}
    \caption{Vision transformer encoding through multiple layers. An image is received, fed through the ViT encoder tower represented here by 9 layers, and finally passed to a projection head that composes the embedding. Vision transformers progressively refine their representation from coarse to detailed. Simple images may reach sufficient quality by the middle layers, which ICAR explores.}
    \label{fig:vit_refinement}
\end{figure*}

Single-modality methods have explored adaptive computation by adjusting processing based on input difficulty. BranchyNet~\cite{teerapittayanon2016branchynet} introduced early exits with classifiers at intermediate layers, enabling confident predictions to exit early while difficult cases continue through all layers. Several methods adapt early exits for Vision Transformers through patch processing, token removal, and routing strategies~\cite{yin2022avit,chen2023cf,rao2021dynamicvit,bolya2023token,riquelme2021scaling}.

These vision-only methods cannot handle vision-language models because they ignore cross-modal alignment requirements. Image-text retrieval needs embeddings from different processing depths to remain compatible in the same semantic space for text matching: a fundamental challenge that existing adaptive methods cannot address.

Limited work exists on adaptive computation for vision-language models. DeeCap~\cite{fei2022deecap}, MuE~\cite{tang2023you}, and FREE~\cite{bajpai2025free} focus on generative settings, while LGViT~\cite{xu2023lgvit} studies early exiting for vision-only classification. For retrieval, HiVLP~\cite{chen2022hivlp} uses hierarchical features but requires re-ranking due to separate embedding spaces. In contrast, ICAR targets single-stage image--text retrieval and enforces embedding compatibility across vision depths so that all image embeddings remain directly comparable to the text embedding (Section~\ref{sec:icar}).

\section{Image Complexity Detection Method}\label{sec:complexity}

\subsection{Proposed Method: ConvNeXt-IC}

\begin{figure}[htbp]
    \centering
    \subfloat[Low Image Complexity]{%
        \begin{minipage}{0.33\textwidth}
            \centering
            \includegraphics[width=\textwidth]{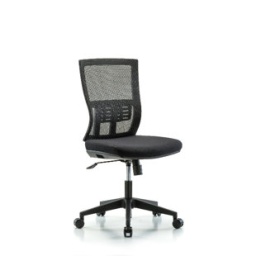}
            \label{fig:complexity_examples:low}
        \end{minipage}
    }
    \hspace{0.05\textwidth}
    \subfloat[High Image Complexity]{%
        \begin{minipage}{0.49\textwidth}
            \centering
            \includegraphics[width=\textwidth]{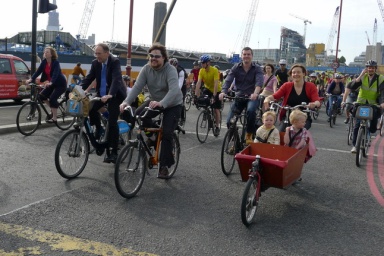}
            \label{fig:complexity_examples:high}
        \end{minipage}
    }
    \caption{Example images from the LAION dataset illustrating complexity variation. (a) Simple images feature single objects with minimal context. (b) Complex images contain multiple objects, spatial relationships, and rich contextual information.}
    \label{fig:complexity_examples}
\end{figure}

ConvNeXt-IC predicts whether an image is simple or complex, providing the routing signal for ICAR. Figure~\ref{fig:complexity_examples} shows representative examples from LAION-5B~\cite{schuhmann2022laion}.

These examples illustrate the range of complexity that ViTs must handle. Existing image complexity assessment methods treat complexity assessment as a representation learning problem, building specialised architectures to capture complexity patterns~\cite{feng2022ic9600,guo2023image,li2025unlocking}. However, the core challenge is discriminating between different levels of visual complexity. We therefore treat complexity assessment as a classification problem, using efficient classification backbones instead of specialised complexity representation architectures.

We therefore fine-tune ConvNeXt-V2-N (224x224) \cite{woo2023convnext}, the smallest of the ConvNeXt-V2 ImageNet-22K~\cite{deng2009imagenet} family of foundation classification models. This choice prioritises efficiency for real-time routing decisions while maintaining the discriminative power needed for accurate complexity assessment. We implement ConvNeXt-IC with two configurations: (1) as a regression model (using a single linear layer predicting continuous scores); and (2) as a binary classification model (using a lightweight two-layer MLP with cross-entropy loss). Both configurations are fine-tuned end-to-end using standard data augmentation (random crops, horizontal flips, colour jittering) on the benchmark datasets.

\subsection{Benchmark Evaluation on IC9600}\label{sec:ic9600}
We benchmark ConvNeXt-IC on IC9600~\cite{feng2022ic9600} using its regression head, a standard dataset for image complexity assessment. IC9600 contains 9,600 images labelled with complexity scores from 1 (simple) to 5 (complex). Performance is measured following the IC9600 evaluation protocol~\cite{feng2022ic9600} using standard regression metrics: Pearson Correlation Coefficient (PCC) and Spearman Rank Correlation Coefficient (SRCC) assess how well predicted scores correlate with human ratings, while Root Mean Square Error (RMSE) measures prediction accuracy. Images per second (img/s) inference speed is also reported to assess computational efficiency. Table~\ref{tab:ic9600_results} compares results to existing methods.

\begin{table}[ht]
\centering
\begin{tabular}{p{1.4in}p{0.6in}p{0.6in}p{0.6in}p{0.6in}} \hline
Method & PCC $\uparrow$ & SRCC $\uparrow$ & RMSE $\downarrow$ & img/s $\uparrow$ \\ \hline
HyperIQA~\cite{su2020blindly} & 0.935 & 0.926 & 0.204 & $\sim$61 \\
CLIPIQA~\cite{wang2022exploring} & 0.898 & 0.897 & 0.078 & $\sim$83 \\
TOPIQ~\cite{chen2024topiq} & 0.944 & 0.938 & 0.049 & $\sim$125 \\
CLICv2~\cite{liu2025clicv2} & 0.933 & 0.927 & - & - \\
ICNet~\cite{feng2022ic9600} & 0.949 & 0.945 & 0.053 & $\sim$397 \\
MICM~\cite{li2025unlocking} & 0.953 & 0.943 & 0.049 & $\sim$0.01 \\
ICCORN~\cite{guo2023image} & 0.955 & 0.951 & - & - \\
ConvNeXt-IC & 0.959 & 0.956 & 0.045 & $\sim$1744 \\ \hline \\ \\
\end{tabular}
\caption{Performance comparison on the IC9600 dataset. $\uparrow$ indicates higher is better and $\downarrow$ indicates lower is better.}
\label{tab:ic9600_results}
\end{table}

ConvNeXt-IC achieves state-of-the-art performance: 0.959 PCC and 0.956 SRCC, outperforming the previous best method ICCORN (0.955 PCC, 0.951 SRCC) while delivering 1,744 img/s throughput. Most notably, ConvNeXt-IC surpasses methods that were specifically architecturally designed for image complexity detection, like ICNet (0.949 PCC) and MICM (0.953 PCC). This demonstrates that strong, correctly chosen backbones which are then fine-tuned can outperform specialised architectures, providing another instance of the current trend in machine learning where data and compute together can outperform sophisticated architectural design~\cite{halevy2009unreasonable,zhai2022scaling}.

\subsection{Evaluation on a LAION Subset}

While IC9600 provides a controlled benchmark dataset, its images are sourced from curated academic datasets. To complement this benchmark, we train and evaluate ConvNeXt-IC on a randomly sampled LAION subset that we manually annotate as `complex' or `simple', reflecting the web distribution where ICAR makes routing decisions. IC9600 serves as the main image complexity prediction evaluation, and this LAION subset complements it by evaluating performance on the web distribution ConvNeXt-IC is intended for.

We sample 3,000 images from LAION-COCO (a subset of LAION-5B)~\cite{schuhmann2022laion} and manually annotate them into complexity classes of `complex' and `simple'. The annotation process naturally resulted in the classes being distributed as 49.5\% simple and 50.5\% complex. Note that, this annotation process was conducted solely by the lead researcher of this paper. Therefore, while ConvNeXt-IC is benchmarked on IC9600 (Section~\ref{sec:ic9600}), results on this LAION subset are best interpreted as an evaluation of performance on the web distribution used for routing.

The 3,000 annotated images are split into 2,100/450/450 (train/val/test) sets using stratified splitting under a fixed random seed (42) for reproducibility. ConvNeXt-IC and the ICNet baseline are trained and evaluated under this split, and results are reported on the held-out test set. For both methods, the prediction threshold is empirically chosen on the validation set and applied to the held-out test set. Table~\ref{tab:complexity_results} shows the comparative performance.

\begin{table}[htbp]
\centering
\begin{tabular}{p{1in}p{0.6in}p{0.6in}p{0.6in}p{0.8in}p{0.6in}} 
\hline 
Method & Precision & Recall & F1 & ROC AUC & img/s \\ \hline
ICNet~\cite{feng2022ic9600} & 0.864 & 0.823 & 0.843 & 0.924 & 397 \\
ConvNeXt-IC & 0.872 & 0.853 & 0.863 & 0.944 & 1744 \\ \hline \\
\end{tabular}
\caption{Evaluation results on the LAION subset (held-out test set) for image complexity assessment.}
\label{tab:complexity_results}
\end{table}

ConvNeXt-IC achieves 0.863 F1-score and 0.944 ROC AUC, outperforming ICNet (0.843 F1, 0.924 ROC AUC) with its 4.4$\times$ efficiency advantage (1,744 vs 397 images/second). These results support the use of ConvNeXt-IC as ICAR's routing module for web images.

\section{Proposed ICAR Architecture}\label{sec:icar}

\subsection{Model Design and Dual-Path Training}

\begin{figure*}[htbp]
    \centering
    \includegraphics[width=0.9\linewidth]{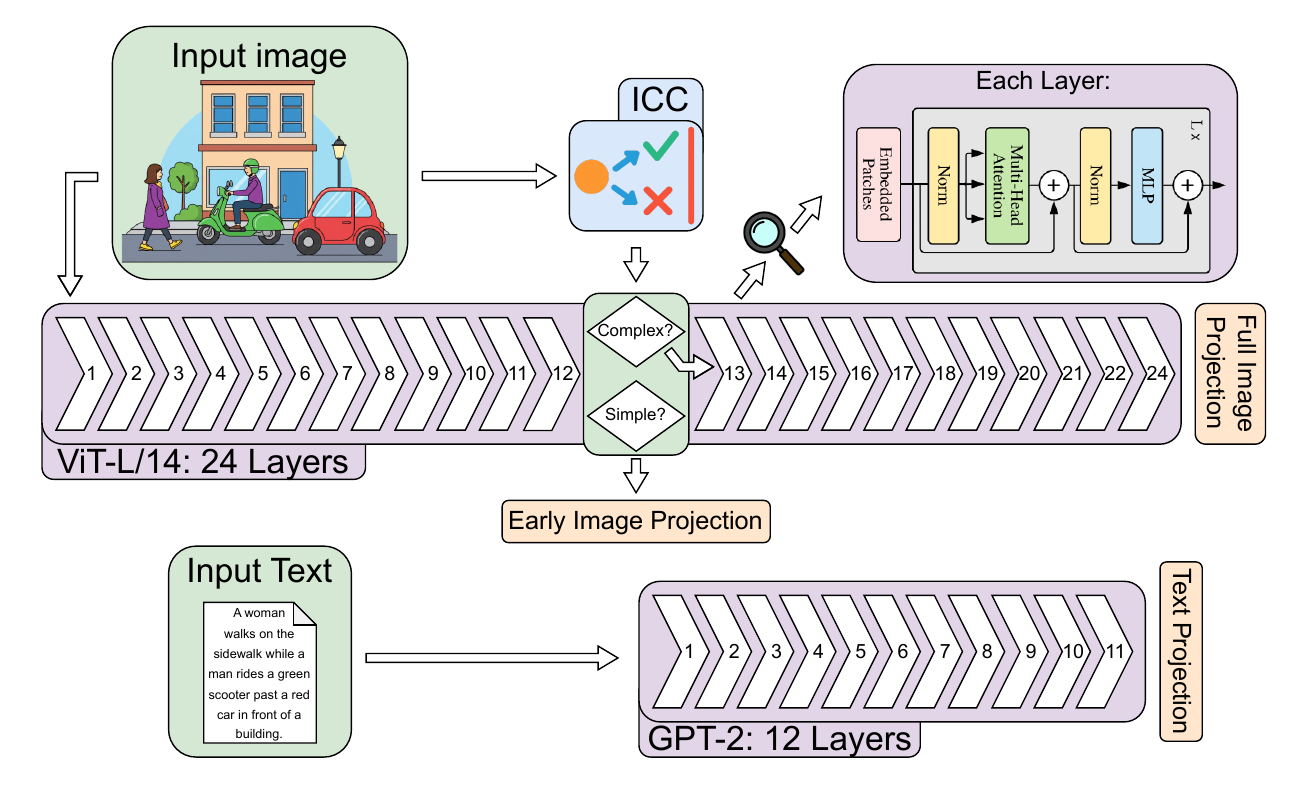}
    \caption{Image Complexity-Aware Retrieval (ICAR). ICC routes simple images to early exits (layers 8, 12, 16, or 20) while complex images use full 24-layer processing. Both paths produce embeddings in a unified space for direct text matching.}
    \label{fig:ICAR}
\end{figure*}

ICAR integrates an Image Complexity Classifier (ICC), powered by ConvNeXt-IC, with adaptive ViT-L/14 processing. Figure~\ref{fig:ICAR} shows the architecture: simple images exit early at layers 8, 12, 16, or 20, while complex images process through all 24 layers. The approach applies dual-path training to produce compatible embeddings from both paths in the same semantic space, addressing the cross-modal alignment challenge in vision-language retrieval.

\textbf{Architecture:} ICAR consists of (1) ICC for routing decisions (powered by ConvNeXt-IC; 2.45 GFLOPs overhead), (2) ViT-L/14 with early exit points, and (3) identical projection heads ensuring embedding compatibility. Early exit at layer 8 saves 44\% computation (108 GFLOPs), while later exits provide efficiency-quality trade-offs. Text processing remains unchanged at 13.3 GFLOPs.

\textbf{Dual-Path Training:} During training, each image-text pair passes through both the complexity-determined early exit path and the full processing path, creating two vision embeddings that must both align with the corresponding text embedding. We optimise a combined loss:

\begin{equation}
\mathcal{L}_{\text{total}} = \alpha \mathcal{L}_{\text{early}} + (1 - \alpha) \mathcal{L}_{\text{full}}
\end{equation}

where $\mathcal{L}_{\text{early}}$ and $\mathcal{L}_{\text{full}}$ are CLIP contrastive losses from respective paths, and $\alpha = 0.5$ weights both paths equally. This creates multi-scale supervision that prevents early layers from over-specialising while ensuring both processing paths learn compatible representations within the same semantic space.

\begin{figure}[htbp]
    \centering
    \includegraphics[width=\linewidth]{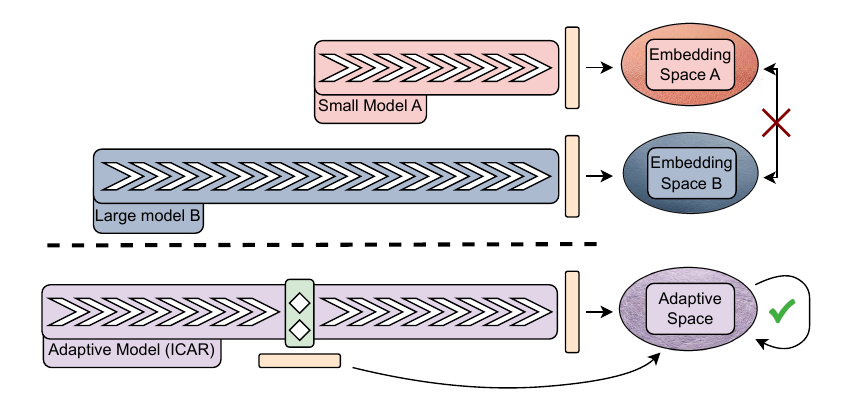}
    \caption{ICAR's unified embedding space enables direct comparison between early-exit and full-processing embeddings for text matching, while reranking approaches require separate models with different embedding spaces and two-stage processing.}
    \label{fig:embeddingspaces}
\end{figure}

This unified approach eliminates the storage overhead and computational complexity of reranking methods, which require separate embedding storage and two-stage retrieval pipelines. All embeddings, regardless of processing depth, remain directly comparable using cosine similarity without additional processing steps.

\subsection{Experimental Setup}

\textbf{Datasets:} MS-COCO~\cite{lin2014microsoft} and Flickr30K~\cite{young2014flickr30k} are standard image--text retrieval benchmarks with curated image-caption pairs. To evaluate retrieval under conditions that are more representative of real-world data, the retrieval pool is augmented using LAION-COCO~\cite{schuhmann2022laion}. LAION~\cite{schuhmann2022laion} is a large-scale dataset of image--text pairs that was collected through automation from publicly available web sources, and LAION-COCO is a subset of it which contains text in the same caption style as MS-COCO and Flickr30K. Training uses the MS-COCO (82K) and Flickr30K (29K) training sets, and evaluation uses the respective test sets (5K test set for MS-COCO and 1K test set for Flickr30K) each augmented with an additional 100K LAION-COCO samples.

\textbf{Variants and baseline:} ICAR-x/24 denotes an early exit after layer x, out of 24 vision layers. Four variants are evaluated with x=8, 12, 16, and 20 (ICAR-8/24, ICAR-12/24, ICAR-16/24, ICAR-20/24), covering compute--quality trade-offs from early to late exits. ViT-L/14 (OpenCLIP) is used as the baseline because ICAR is built directly on top of the same backbone~\cite{ilharco_gabriel_2021_5143773}. This isolates the effect of ICC-based routing and early exits.

\textbf{Training:} ICAR is fine-tuned from OpenCLIP ViT-L/14~\cite{ilharco_gabriel_2021_5143773} using AdamW \cite{loshchilov2019decoupled} with differential learning rates (1e-6 backbone, 1e-4 projection heads) for 10 epochs. Weight decay is set to 0.01 with gradient clipping (1.0), cosine annealing with warmup, and equal weighting of early and full losses ($\alpha=0.5$). Training uses a batch size of 128, and the best validation checkpoint across epochs is used for evaluation.

\textbf{Tasks and metrics:} We evaluate text-to-image retrieval under two settings. For instance-level retrieval, each query has a single correct image, and retrieval quality is measured using Recall@k (R@k). For category-level retrieval, we map samples to MS-COCO semantic categories and treat retrievals as correct when the query and retrieved image belong to the same category, and retrieval quality is measured using mean Average Precision@k (mAP@k). GFLOPs are reported as a weighted average over the routing distribution (49.5\% simple, 50.5\% complex), accounting for the chosen early-exit depth and ICC overhead. Lower GFLOPs indicate less compute per sample and are therefore preferable for efficiency.

\pagebreak

\section{Results and Analysis}\label{sec:results}

\subsection{Retrieval Performance and Key Insights}

Tables~\ref{tab:instance_results} and~\ref{tab:category_results} show the trade-off between early exiting and retrieval quality across instance-level and category-level retrieval.

\begin{table}[htbp]
\centering
\begin{tabular}{p{0.9in}>{\centering\arraybackslash}p{0.4in}>{\centering\arraybackslash}p{0.4in}>{\centering\arraybackslash}p{0.4in}@{\hspace{0.12in}}>{\centering\arraybackslash}p{0.4in}>{\centering\arraybackslash}p{0.4in}>{\centering\arraybackslash}p{0.4in}@{\hspace{0.12in}}>{\centering\arraybackslash}p{0.5in}} \hline
\multicolumn{8}{c}{Instance Retrieval (R@k)}  \\ \hline
& \multicolumn{3}{c}{Flickr + LAION} & \multicolumn{3}{c}{COCO + LAION} & \\
& @1 & @5 & @10 & @1 & @5 & @10 & GFLOPs \\ \hline
ICAR-20/24 & 71.1 & 89.8 & 93.7 & 47.4 & 72.7 & 81.1 & 164.41 \\
ICAR-16/24 & 69.5 & 88.9 & 93.6 & 47.1 & 72.2 & 81.2 & 151.05 \\
ICAR-12/24 & 68.5 & 88.8 & 92.9 & 45.1 & 70.2 & 79.3 & 137.68 \\
ICAR-8/24 & 67.4 & 87.6 & 92.2 & 42.9 & 68.3 & 77.4 & 124.31 \\ \hline
$\text{ViT-L/14}_{\text{FT}}$~\cite{ilharco_gabriel_2021_5143773} & 71.0 & 89.9 & 93.8 & 48.5 & 73.5 & 81.8 & 175.33 \\ \hline \\ 
\end{tabular}
\caption{Instance-level text-to-image retrieval results, measured using R@k. The fine-tuned (FT) ViT-L/14, the architecture which the proposed ICAR is based on, is used as a baseline.}
\label{tab:instance_results}
\end{table}

\begin{table}[htbp]
\centering
\begin{tabular}{p{0.9in}>{\centering\arraybackslash}p{0.4in}>{\centering\arraybackslash}p{0.4in}>{\centering\arraybackslash}p{0.4in}@{\hspace{0.12in}}>{\centering\arraybackslash}p{0.4in}>{\centering\arraybackslash}p{0.4in}>{\centering\arraybackslash}p{0.4in}@{\hspace{0.12in}}>{\centering\arraybackslash}p{0.5in}} \hline 
\multicolumn{8}{c}{Category Retrieval (mAP@k)} \\ \hline
& \multicolumn{3}{c}{Flickr + LAION} & \multicolumn{3}{c}{COCO + LAION} & \\
& @10 & @50 & @100 & @10 & @50 & @100 & GFLOPs \\ \hline
ICAR-20/24 & 83.8 & 71.7 & 64.2 & 92.9 & 87.6 & 84.7 & 164.41 \\
ICAR-16/24 & 83.4 & 71.4 & 64.0 & 93.1 & 87.8 & 84.8 & 151.05 \\
ICAR-12/24 & 83.4 & 70.5 & 63.4 & 92.8 & 87.4 & 84.4 & 137.68 \\
ICAR-8/24 & 82.2 & 69.7 & 62.7 & 93.1 & 87.9 & 84.9 & 124.31 \\ \hline
$\text{ViT-L/14}_{\text{FT}}$~\cite{ilharco_gabriel_2021_5143773} & 84.3 & 70.2 & 64.1 & 92.6 & 86.2 & 82.2 & 175.33 \\ \hline \\
\end{tabular}
\caption{Category-level text-to-image retrieval results, measured using mAP@k.}
\label{tab:category_results}
\end{table}

\textbf{Instance-level retrieval:} Instance-level retrieval evaluates exact matching between specific image--text pairs, which places more emphasis on fine-grained details. In our setup, the additional LAION-COCO samples primarily act as distractor candidates for these queries, so routing simple candidates through earlier exits mainly affects efficiency, and only harms quality if early-exit embeddings become insufficiently discriminative. Overall, instance-level retrieval shows a gradual performance decline as the exit layer becomes earlier; for ICAR-8/24, the aggregate recall scores across Flickr+\allowbreak LAION and COCO+\allowbreak LAION are 95.0\% of the baseline\footnote{Performance retention calculated using aggregate RSUM~\cite{li2022image} methodology: sum of all ICAR-8/24 retrieval scores divided by sum of baseline scores across both datasets.}.

\textbf{Category-level retrieval:} Category-level retrieval treats any candidate from the same category as relevant, so simple web images can become valid matches for broader queries. This makes category-level performance a direct test of whether early-exit embeddings preserve coarse semantic information well enough for matching. In Table~\ref{tab:category_results}, ICAR variants generally maintain category-level performance with small, dataset-specific differences: on COCO+\allowbreak LAION, mAP@10 can be marginally higher (e.g., 93.1 vs 92.6 baseline), whereas on Flickr+\allowbreak LAION, ICAR variants remain within \(\pm\)1--2 points of baseline across mAP@k.

\textbf{Coarse-to-fine features:} The results support our hypothesis about the progressive nature of transformer processing: early layers capture coarse semantic features that are often sufficient for category matching, while deeper layers refine instance-level discriminative details. The strong category-level performance is consistent with dual-path training providing multi-scale supervision that acts as implicit regularisation, improving robustness of early-exit embeddings for coarse matching.

\subsection{Efficiency Analysis and Impact}

\textbf{Routing coverage:} Most early exits come from the LAION-COCO portion of the retrieval pool rather than the benchmark test images. ICC routes 195/5,000 MS-COCO test images (3.9\%) and 2/1,000 Flickr30K test images (0.2\%) to early exits, but routes 49,314/100,000 LAION-COCO candidates (49.3\%). This indicates that most compute savings come from routing web candidates, while curated benchmark images are rarely routed to early exits. This behaviour is desirable for our target use case: curated benchmarks tend to contain more complex images, whereas web data includes a broader mix of simple and complex samples. ICAR therefore targets the part of the retrieval pool where there is the most opportunity for compute savings.

\begin{figure}[htbp]
    \centering
    \begin{minipage}{0.48\textwidth}
        \centering
        \includegraphics[width=\textwidth]{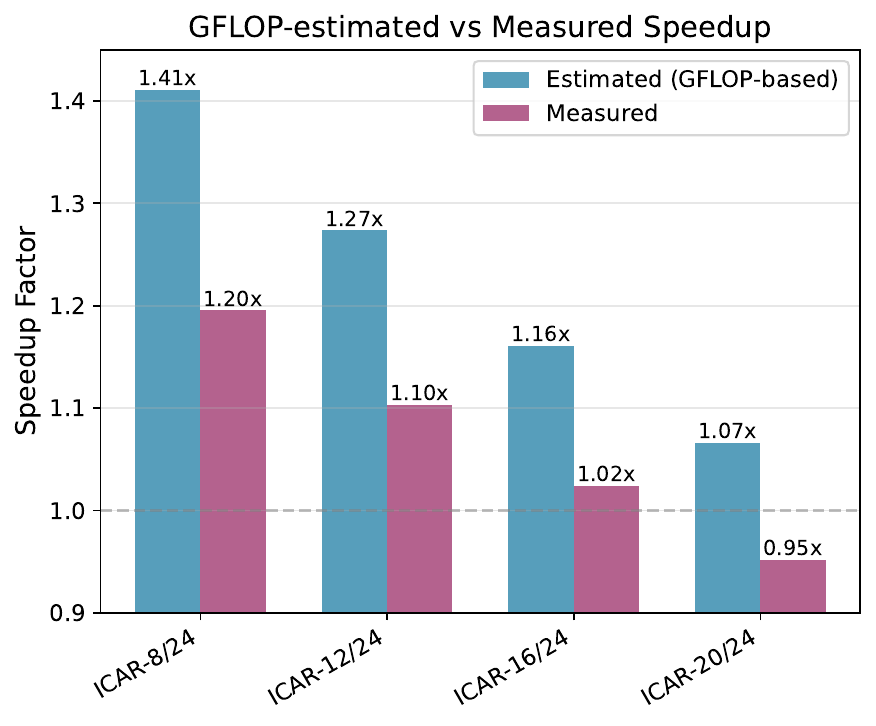}
    \end{minipage}
    \hfill
    \begin{minipage}{0.48\textwidth}
        \centering
        \includegraphics[width=\textwidth]{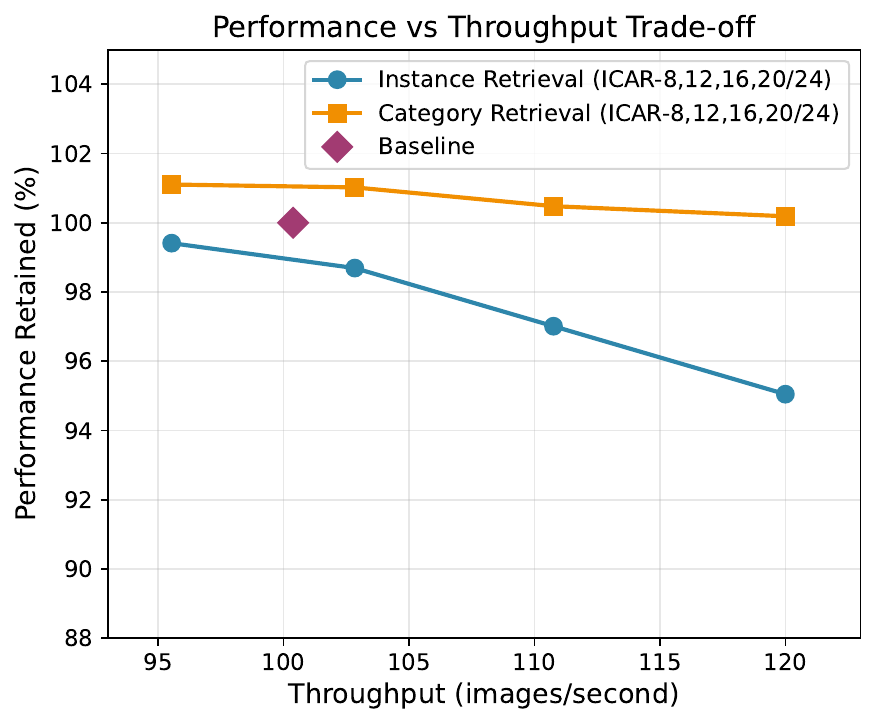}
    \end{minipage}
\caption{ICAR performance analysis. (a) Estimated image encoding speedups based on GFLOP reduction versus measured image encoding speedups on an NVIDIA A100 GPU. (b) Performance retention aggregated from Tables~\ref{tab:instance_results} and~\ref{tab:category_results} showing different performance trends on instance versus category retrieval on Flickr30K+LAION-COCO and MS-COCO+LAION-COCO.}
    \label{fig:icar-performance}
\end{figure}

\textbf{Measured vs estimated speedups:} ICAR achieves 20\% faster image encoding (ICAR-8/24: 1.195$\times$, 120.0 vs 100.4 images/second) with GFLOP-estimated potential up to 1.41$\times$. The gap between estimated and measured speedups reflects additional overheads beyond raw transformer compute, including routing, framework and kernel overheads, so GFLOP reduction is an upper bound on throughput gains. Figure~\ref{fig:icar-performance}(a) compares estimated versus measured A100 speedups, while Figure~\ref{fig:icar-performance}(b) shows differential impact: category-level retrieval maintains or exceeds baseline (ICAR-8/24: 100.1\% retention), whereas instance-level shows expected degradation.

\begingroup
\interfootnotelinepenalty=10000
\textbf{Production-scale impact:} If efficiency improvements such as ICAR's 20\% were implemented at big tech production scale, processing Google Photos' 6 billion daily images~\cite{GoogleIO2024} could save 667 A100 GPU-hours daily and 133,640 kWh annually, equivalent to eliminating 16.6 metric tonnes of CO2 emissions\footnote{Projection (A100): Google Photos processes \(6\times 10^9\) images/day~\cite{GoogleIO2024}. Assume 500 img/s per GPU~\cite{NVIDIADocsOptimization}. GPU-hours/day \(= \frac{6\times 10^9}{500 \cdot 3600} = 3{,}333\). Annual energy \(= 3{,}333 \cdot 0.5\,\text{kW} \cdot 365 \cdot 1.10 = 668{,}002\)\,kWh (PUE 1.10)~\cite{GoogleDataCenters}. A 20\% reduction saves 667 GPU-hours/day and 133{,}640\,kWh/year. CO\(_2\) uses UK grid intensity~\cite{CarbonBrief2024} (\(\approx 16.6\) t/year).}.
\endgroup

\section{Conclusion}\label{sec:conclusion}

This paper presented ICAR (Image Complexity-Aware Retrieval), the first adaptive computation method for image-text retrieval that addresses the cross-modal alignment challenge through unified embedding spaces. At the core of ICAR is the ICC module, a lightweight image complexity classifier powered by ConvNeXt-IC which predicts whether an image is simple or complex and routes simple images to early exits while complex images use full processing. During training, both the early-exit and full paths are jointly supervised via dual-path contrastive learning so that all vision embeddings remain compatible with the text embedding in a single unified space. By recognising complexity assessment as a classification task, ConvNeXt-IC achieves state-of-the-art performance (0.959 PCC and 0.956 SRCC, 4.4$\times$ faster complexity prediction), and ICAR's dual-path training enables 20\% faster image encoding with maintained retrieval quality.

Future work could enhance the approach through dynamic exit selection based on intermediate confidence scores, end-to-end integration of complexity assessment within the vision-language learning objective, and production-grade optimisations to bridge the gap between GFLOP-estimated (1.41$\times$) and measured (1.195$\times$) image encoding speedups. Additional work includes evaluating other backbones and patch sizes, and studying routing calibration under distribution shift.

\begin{credits}
{\renewcommand{\ackname}{Data and code availability}%
\subsubsection{\ackname}
To facilitate reproducibility and future research, all code, trained model weights and data are publicly available at \url{https://github.com/MikelWL/ICAR}.
}

\subsubsection{\discintname}
The authors have no competing interests to declare that are relevant to the content of this article.
\end{credits}

\bibliographystyle{splncs04}
\bibliography{references}

\end{document}